# Solar System Formation Deduced from Observations of Matter

J. Marvin Herndon

Transdyne Corporation
San Diego, California 92131 USA

August 9, 2004

## Abstract

Aspects of our Solar System's formation are deduced from observations of the chemical nature of matter. Massive cores are indicative of terrestrial-planet-composition-similarity to enstatite chondrite meteorites, whose highly-reduced state of oxidation may be thermodynamically stable in solar matter only at elevated temperatures and pressures. Consistent with the formation of Earth as envisioned by Arnold Eucken, thermodynamic considerations lead to the deduction that the terrestrial planets formed by liquid-condensation, raining out from the central regions of hot, gaseous protoplanets. The mass of protoplanetary-Earth, estimated to be 275-305$m_E$, is similar to the mass of Jupiter, 318$m_E$. Solar primordial gases and volatile elements were separated from the terrestrial planets early after planet formation, presumably during some super-luminous solar event, perhaps even before Mercury had completely formed. The pre-super-luminosity-terrestrial-planet mass distribution appears to be more consistent with observations of the close-in gas-giants of other planetary systems than the present-day-terrestrial-planet mass distribution. Although primarily formed by raining out from the centers of hot, gaseous protoplanets, evidence for the Earth and, by inference for the other terrestrial planets, suggests some outer, minor, secondary accretion of oxidized matter in the grain-growth accumulation way envisioned by George W. Wetherill. By contrast, the so-called "standard model" for our Solar System's formation has flawed conceptual underpinnings and would lead to the contradiction of terrestrial planets having insufficiently massive cores.

## Introduction

Observations over the past decade of hundreds of planetary systems provide rich opportunities to address the question of how special is our own Solar System (Beer et al. 2004). But, in attempting to reconcile observations of other planetary systems, it is important to consider *observations*, not simply models, of our own. As recently documented (Herndon 2004a), the Earth science literature is turgid with decades of reports of investigations that fail to follow long-established, ethical standards of science. One major problem is that many investigators contrive models that often ignore contradictory scientific evidence and that are based upon arbitrary assumptions, or based upon other models, which are themselves based upon arbitrary assumptions; there is spill-over into the underpinnings of the so-called "standard model" of solar system formation.

I hold that the purpose of science is not to make arbitrary models based upon assumptions, but rather, to determine the true nature of the Earth and the Cosmos, which can be done by making fundamental discoveries and by discovering fundamental quantitative relationships in nature. In the following, I make use of *observations* of the chemical nature of the matter that comprises our Solar System. I present from a historical perspective a brief outline of the development of the "standard model" of solar system formation in connection to that matter and reveal the flawed underpinnings of that "standard model". I then make a few generalizations that might be useful for understanding somewhat better the development of our own Solar System and for making comparisons to other planetary systems.

## Basis of Misunderstanding

According to Beer et al. (2004) "This standard model for the solar system (Mizuno 1980; Stevenson 1982; Wetherill 1980) assumes that planets form initially through the agglomeration of dust into grains, pebbles, rocks and thence planetesimals within a gaseous disc, that these planetesimals coalesce to form planetary cores, and that finally (for the giant planets) these cores use gravity to accrete gas from the ever-present disc." But to understand the underlying flaw one must look a bit further back in time.

In 1944, on the basis of thermodynamic considerations, Arnold Eucken suggested core-formation in the Earth, as a consequence of successive condensation from solar matter, on the basis of relative volatility, from the central region of a hot, gaseous protoplanet, with molten iron metal first raining out at the center (Eucken 1944). For a time hot, gaseous protoplanets were discussed (Kuiper 1951a; Kuiper 1951b; Urey 1952), but emphasis changed abruptly with the publication by Cameron (1963) of his diffuse solar nebula models at pressures of about $10^{-5}$ bar.

During the late 1960s and early 1970s, the so-called "equilibrium condensation" model was contrived and widely promulgated (Larimer & Anders 1970). That model was predicated upon the assumption that the mineral assemblage characteristic of ordinary chondrite meteorites (Table 1) formed as condensate from a gas of solar composition at pressures of about $10^{-5}$ bar. At the time scientists almost universally believed that the Earth is like an ordinary chondrite meteorite. Consequently, the idea that dust agglomerated into grains, pebbles, rocks and then into planetesimals seemed, superficially at least, to explain planet formation and planet composition. The problem, however, is that the "standard model" of solar system formation would result in the terrestrial planets having insufficiently massive cores, a profound contradiction to what is observed. The reason, as discussed below, is two-fold: The "equilibrium condensation" model is *wrong* and the Earth in its composition is *not* like an ordinary chondrite meteorite.

## Nature of Solar System Matter

The constancy in isotopic compositions of most of the elements of the Earth, the Moon, and the meteorites indicates formation from primordial matter of common origin. Primordial elemental composition is yet manifest and determinable to a great extent in the photosphere of the Sun. The less volatile rock-forming elements, present in the outer regions of the Sun, occur in nearly the same relative proportions as in chondritic meteorites. But chondrites differ from one another in their respective proportions of major elements, in their states of oxidation, mineral assemblages, and oxygen isotopic compositions and, accordingly, are grouped into three distinct classes: *enstatite*, *carbonaceous* and *ordinary*, as shown in Table 1.

Only five elements (Fe, Mg, Si, S and O) constitute about 95% of the mass of each chondrite meteorite and, by implication, about 95% of the mass of each of the terrestrial planets. Four of those elements (Fe, Mg, Si and S) occur in chondrites in about the same relative proportion as they occur in the photosphere of the Sun, to within a factor of two. Oxygen is the exception, being about 8 times more abundant in the photosphere of the Sun than the sum of the other four elements of either chondrite shown in Table 2. The high relative abundance of oxygen in solar matter poses a serious limitation on the nature of primordial condensates from that medium.

Hans Suess and I demonstrated from thermodynamic considerations that the oxidized iron content of the silicates of ordinary chondrites is *inconsistent* with formation from solar matter, as purported by the "equilibrium condensation" model, and instead is indicative of their formation from a gas phase depleted in hydrogen by a factor of about 1000 relative to solar composition (Herndon & Suess 1977). Subsequently, I showed that oxygen depletion, relative to solar matter, was also required, otherwise essentially all of the elements would be observed combined with oxygen as they are in the hydrous C1 carbonaceous chondrites (Table 1). I also showed that if the mineral assemblage characteristic of ordinary chondrites could exist in equilibrium with a gas of solar composition, it is at most only at a single low temperature, if at all (Herndon 1978). Such a mineral assemblage, therefore, cannot legitimately be assumed to be a primary Solar System condensate. Instead, the ordinary chondrite meteorites appear to have formed from a mixture of two components, re-evaporated after separation from solar gases, one component being an oxidized primitive matter like C1 chondrites, the other being a partially differentiated planetary component from enstatite-chondrite-like matter (Herndon 2004b).

The oxidation states of chondritic matter and, by implication, the oxidation states of the terrestrial planets are set by the nature of those condensates and the circumstances of the separation of those condensates from their primordial gases. After having been separated from the solar gases, the oxidation state of the condensate cannot be appreciably changed.

The relative mass of the iron alloy component, which forms the cores of the terrestrial planets, is an inverse function of the oxygen content of that condensate (Herndon & Suess 1977). Condensation from a cooling gas of solar composition at $10^{-5}$ bar is expected from thermodynamic considerations to lead to an oxygen-rich condensate not unlike the hydrous C1 carbonaceous chondrites (Table 1). After separation from solar gases, such an oxidized condensate, if melted or re-evaporated and re-condensed, will lead to condensed matter not unlike the oxygen-rich anhydrous carbonaceous chondrites (Table 1), which, if agglomerated into successive larger sized objects, in the manner of the "standard model" of solar system formation, would result in the terrestrial planets having insufficiently massive cores; a circumstance that is contrary to all current observations of those planets.

For decades there has been a wide-spread misconception that the Earth is like an ordinary chondrite meteorite (Herndon 2004a). After realizing that the composition of the Earth's inner core was not necessarily partially crystallized iron (Herndon 1979), I related the parts of the Earth by fundamental mass ratios to the parts of a particular enstatite chondrite (Herndon 1980; Herndon 1982; Herndon 1996). It is easily demonstrable that the chondritic-Earth is in the main like an enstatite chondrite and not like an ordinary chondrite.

Imagine melting a chondrite in a gravitational field. The iron metal and iron sulfide components will alloy together, forming a dense liquid that will settle beneath the silicates like steel on a steel-hearth. The Earth is like a spherical steel-hearth with a fluid iron alloy core surrounded by a silicate mantle. The Earth's core comprises about 32.5% by mass of the Earth as a whole. Observe the percentage by mass of the alloy portion of each ordinary and enstatite chondrite for which data are available, as shown in Figure 1 from Herndon (2004a). Note that, if the Earth has a chondritic composition, as widely believed for good reason, then the Earth is in the main like an enstatite chondrite and *not* like an ordinary chondrite.

Enstatite chondritic matter is the most highly-reduced (oxygen-poor), naturally occurring mineral assemblage known and, strangely, is generally ignored in the so-called "standard model" of solar system formation. But that mineral assemblage exists, and should not be ignored as understanding the nature of its origins will shed much light on the origin of our Solar System.

The formation of enstatite chondrites has posed something of an enigma for those who contrive models, such as the "standard model", because, at pressures of about $10^{-5}$ bar, solar matter is much too oxidizing. Unlike the hydrous C1 carbonaceous chondrites, which consist of low-temperature minerals, the E4 chondrites, such as the Abee enstatite chondrite, show evidence of their components having been at melt or near-melt temperatures. Significantly, each of these strikingly different meteorites has virtually the same trace element content, an indication of a relatively simple chemical history.

On the basis of thermodynamic considerations, Hans Suess and I showed that some of the minerals of enstatite meteorites could form at high temperatures in a gas of solar composition at pressures above about 1 bar, provided that thermodynamic equilibria are frozen in at near-formation temperatures (Herndon & Suess 1976). At such pressures, molten iron, together with the elements that dissolve in it, is the most refractory condensate. Suess and I did not propose a specific model for the formation of enstatite chondrites and, indeed, there is much to verify and learn about the process of condensation from near the triple point of solar matter, but the glimpses we have seen are remarkably similar to vision of Euchen (1944), *i.e*., molten iron raining out in the center of a hot, gaseous protoplanet.

## Deductions from Observations of Matter

There seems to be little doubt that the oxidized, hydrous C1 chondrites originate in the outer reaches of our Solar System, regions sufficiently cold to permit the retention of water in the vacuum of space for billions of years. The oxidation state of C1 chondrites is just what one would expect for their having been derived from solar-matter low-pressure condensation to low temperatures.

There are reasons to associate the highly reduced matter of enstatite chondrites with the inner regions of the Solar System: (1) The regolith of Mercury appears from reflectance spectrophotometric investigations (Vilas 1985) to be virtually devoid of FeO, like the silicates of the enstatite chondrites (and unlike the silicates of other types of chondrites); (2) E-type asteroids (on the basis of reflectance spectra, polarization, and albedo), the presumed source of enstatite meteorites, are, radially from the Sun, the inner most of the asteroids (Zellner et al. 1977); (3) Only the enstatite chondrites and related enstatite achondrites have oxygen isotopic compositions indistinguishable from those of the Earth and the Moon (Clayton 1993); and, (4) Fundamental mass ratios of major parts of the Earth (geophysically determined) are virtually identical to corresponding (mineralogically determined) parts of certain enstatite chondrites, especially the Abee enstatite chondrite (Herndon 1980; Herndon 1993; Herndon 1996).

In the absence of evidence to the contrary, the observed enstatite-chondritic composition of the terrestrial planets permits the deduction that these planets formed by raining out from the central regions of hot, gaseous protoplanets. Estimates of the masses of those protoplanets may be calculated from solar abundance data (Anders & Grevesse 1989) by adding to the condensable, planetary elements their proportionate amount of solar elements that are typically gases (*e.g*., H, He) or form volatile compounds (*e.g*., O, C, N). The mass of protoplanetary-Earth, calculated in that manner, is about 275 to 305 times the mass of the present-day Earth, a value which is quite similar to Jupiter's mass, $318m_E$.

If the gases and ices were magically stripped from Jupiter, there is reason to think that the remains would be an Earth-like mass of predominantly enstatite-chondritic matter, differentiated like the interior of the Earth.

Oxidation state is the fundamental difference between the Earth being like an ordinary chondrite, as previously thought, and really being, as I have shown, like an enstatite chondrite. The oxidation state determines, not only the relative mass of the core, but the elements the core contains. The main consequence of that highly-reduced oxidation state is the existence of certain otherwise lithophile elements in the Earth's core, including uranium. I have demonstrated the feasibility of a nuclear reactor – the georeactor - at the center of the Earth as the energy source for the geomagnetic field (Herndon 1993; Herndon 1994; Herndon 1996). Numerical simulations conducted at Oak Ridge National Laboratory have confirmed that the georeactor would operate as a fast neutron breeder reactor, and could operate over the lifetime of the Earth, and would produce $^{3}$He and $^{4}$He in the same range of ratios as observed in helium escaping from the Earth's interior, which is strong evidence, indeed (Herndon 2003; Hollenbach & Herndon 2001). Ultimately, antineutrino investigations may provide additional confirmation (de Meijer et al. 2004; Domogatski et al. 2004; Raghavan 2002).

I have also demonstrated the feasibility of planetocentric nuclear reactors as the internal energy sources for the giant planets (Herndon 1992; Herndon 1994). The existence of a central Jovian nuclear reactor would imply uranium in Jupiter's core which would imply an oxidation state like that of an enstatite chondrite and like that of the interior of the Earth. Thus, it may well be that the single, major difference between Jupiter and the terrestrial planets is simply the presence or absence of solar gases and volatile compounds.

The nature of the matter comprising the terrestrial planets suggests that these formed principally by raining out of the centers of giant gaseous protoplanets and might have remained as Jupiter-like gaseous planets were it not that the gases and volatile elements were separated soon after planet formation, presumably early during some super-luminous event, such as a T-Tauri state of deuterium burning associated with the thermonuclear ignition of the Sun. Indeed, there is some reason to think that Mercury was only partially formed at the time of super-luminosity.

Major element fractionation among chondrites has been discussed for decades as ratios relative to Si or Mg. By expressing those elements as ratios relative to Fe, I discovered a new relationship admitting the possibility that ordinary chondrite meteorites are derived from two components: one is a relatively undifferentiated, *primitive* component, oxidized like the C1 chondrites; the other is a somewhat differentiated, *planetary* component, with oxidation state like the highly-reduced enstatite chondrites. I have suggested that the *planetary* component is derived from Mercury's missing elements (Herndon 2004b). If that interpretation is correct, the Fe content of the *planetary* component, only about two-thirds depleted, indicates that Mercury was only partially formed at the time when a fraction of its elements went missing.

From terrestrial seismic data (Dziewonski & Anderson 1981), the gross features of the inner 82% of the Earth, the endo-Earth, *i.e.*, the core-plus-lower mantle, appear relatively simple, consistent with the identification of that part being like an enstatite chondrite and having originated by raining out from the center of a giant gaseous protoplanet. The Earth's upper mantle, on the other hand, displays several seismic discontinuities suggestive of layering. The latter observation, together with observations of nickel, oxidized iron, and an undifferentiated chondritic component in rocks from the upper mantle( Jagoutz et al. 1979), implies that one or more minor, oxidized, components were subsequently added, presumably from a secondary accretion by grain-growth accumulation in the way envisioned by Wetherill (1980).

## Summary and Conclusions

Comparison of other planetary systems with our own should be on the basis of observations, not models based upon assumptions. The underpinnings of the "standard model" of solar system formation are wrong, specifically the "equilibrium condensation" model and the idea of the Earth being like an ordinary chondrite. Because the oxidation state of solar system matter is fixed by the circumstances of its condensation and separation from solar gases and volatile compounds, the models underlying the "standard model" would lead to the contradiction of terrestrial planets having insufficiently massive cores.

The following elements of our Solar System's formation are deduced from observations of the chemical nature of matter: The terrestrial planets are like the highly-reduced enstatite chondrite meteorites. Thermodynamic considerations are consistent with the concept of Eucken (1944) indicating that the terrestrial planets, like the Earth, rained out from the central regions of hot, gaseous protoplanets. From solar abundances, the mass of protoplanetary-Earth was 275-305$m_E$, not very different from the mass of Jupiter, 318$m_E$. Solar primordial gases and volatile elements were separated from the terrestrial planets early after planet formation, presumably during some super-luminous solar event, perhaps even before Mercury had completely formed. Although the terrestrial planets appear to have rained out from the central regions of hot, gaseous protoplanets, evidence suggests some outer, minor, secondary accretion of oxidized matter in the grain-growth accumulation way envisioned by Wetherill (1980).

In comparing other planetary systems with our own, it is now important to ask whether comparisons should be based entirely upon our Solar System's present mass distribution or perhaps we should consider a pre-super-luminosity, terrestrial-gas-giant distribution. In the latter case, that would make our Solar System appear less anomalous to other planetary systems in the representation of Beer et al. (2004). But would that mean that the other observed planetary systems with close-in gas giants have not yet experienced a super-luminosity event, or had such an event occurred *before* the formation of their close-in gas giants, or perhaps they never have or will experience such an event? And how biased are our astronomical observations, being limited by current image resolution, so that we cannot presently discern planetary systems with close-in rocky planets which are the residues from such super-luminosity events, unless those systems have an out-lying gas-giant like Jupiter? These are the questions that astronomers should address, ideally from the standpoint of *observations*, and not from models based upon arbitrary assumptions.

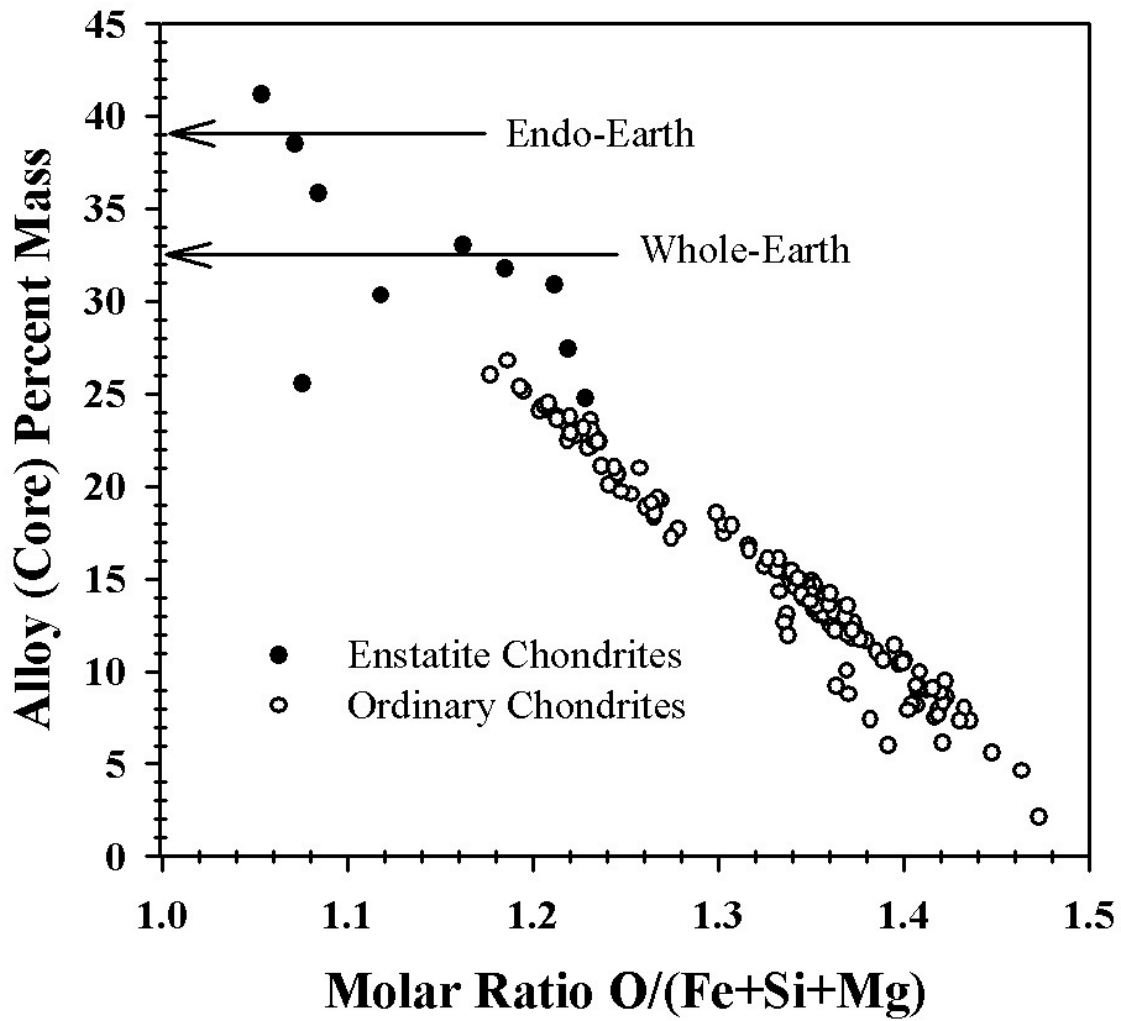

Figure 1. The percent mass of the alloy component of each of 9 enstatite chondrites and 157 ordinary chondrites. This figure clearly shows that, if the Earth is chondritic in composition, the Earth as a whole, and especially the endo-Earth (core-plus-lower mantle), is like an enstatite chondrite and *not* like an ordinary chondrite. The reason is clear from the abscissa which shows the molar ratio of oxygen to the three major elements with which it combines in enstatite chondrites and in ordinary chondrites. This figure also clearly shows that, if the Earth is chondritic in composition, the Earth as a whole, and especially the endo-Earth, has a state of oxidation like an enstatite chondrite and *not* like an ordinary chondrite. From Herndon (2004a).

**Table 1**. The mineral assemblages characteristic of chondritic meteorites. The hydrous C1 carbonaceous chondrites have a state of oxidation characteristic of low-pressure condensation to low temperatures. The highly-reduced enstatite chondrites are similar to the matter of the endo-Earth, the inner 82% of the Earth.

*HYDROUS CHONDRITES*

| | |
|---|---|
| **Carbonaceous Chondrites** | complex hydrous layer lattice silicate  *e.g.* $(Mg, Fe)_6Si_4O_{10}(O, OH)_8$  epsomite, $MgSO_4 \cdot 7H_2O$  magnetite, $Fe_3O_4$ |

*ANHYDROUS CHONDRITES*

| | |
|---|---|
| **Carbonaceous Chondrites** | olivine, $(Fe, Mg)_2SiO_4$  pyroxene, $(Fe, Mg)SiO_3$  pentlandite, $(Fe, Ni)_9S_8$  troilite, FeS |
| **Ordinary Chondrites** | olivine, $(Fe, Mg)_2SiO_4$  pyroxene, $(Fe, Mg)SiO_3$  troilite, FeS  metal, (Fe-Ni alloy) |
| **Enstatite Chondrites** | pyroxene, $MgSiO_3$  complex mixed sulfides    e.g. (Ca, Mg, Mn, Fe)S  metal, (Fe, Ni, Si alloy)  nickel silicide, $Ni_2Si$ |

**Table 2**. Comparison of molar (atom) abundance ratios, normalized to Fe, of the five major elements in the highly-oxidized Orgueil C1 carbonaceous chondrite with those of the highly-reduced Abee enstatite chondrite and with corresponding elements in the photosphere of the Sun (Anders & Grevesse 1989; Baedecker & Wasson 1975). From Herndon (2004a).

| Element Ratio | Orgueil Chondrite | Abee Chondrite | Sun |
| --- | --- | --- | --- |
| Fe/Fe | 1.00 | 1.00 | 1.00 |
| Si/Fe | 1.08 | 1.11 | 1.20 |
| Mg/Fe | 1.19 | 0.82 | 1.12 |
| S/Fe | 0.56 | 0.34 | 0.47 |
| O/Fe | 3.18 | 2.86 | 31.3 |